\documentclass[aps,pra,twocolumn,superscriptaddress]{revtex4-2}

\usepackage{amsmath}
\usepackage{amssymb}
\usepackage{graphicx}
\usepackage{dcolumn}
\usepackage{bm}
\usepackage{calrsfs}  
\usepackage{float}
\usepackage{xcolor}
\usepackage[normalem]{ulem}
\DeclareMathAlphabet{\pazocal}{OMS}{zplm}{m}{n}
\DeclareMathAlphabet\mathbfcal{OMS}{cmsy}{b}{n}
\newcommand{\x}{\mathop{\textbf{X}_{lm} (\hat{\textbf{r}})}}

\begin{document}

\title{Manipulating the photonic Hall effect with hybrid Mie-exciton resonances}

\author{P. Elli Stamatopoulou}
\affiliation{Center for Nano Optics, University of Southern Denmark, Campusvej 55, DK-5230 Odense M, Denmark}
\affiliation{Department of Physics, National Technical University of Athens, GR-15780 Athens, Greece}
\author{Vassilios Yannopapas}
\affiliation{Department of Physics, National Technical University of Athens, GR-15780 Athens, Greece}
%
%
\author{N. Asger Mortensen}
\affiliation{Center for Nano Optics, University of Southern Denmark, Campusvej 55, DK-5230 Odense M, Denmark}
\affiliation{Danish Institute for Advanced Study, University of Southern Denmark, Campusvej 55, DK-5230 Odense M, Denmark}
\author{Christos Tserkezis}
\email{ct@mci.sdu.dk}
\affiliation{Center for Nano Optics, University of Southern Denmark, Campusvej 55, DK-5230 Odense M, Denmark}

\date{\today}

\begin{abstract}
We examine the far-field optical response, under-plane wave excitation in the presence of a static magnetic field, of core-shell nanoparticles involving a gyroelectric component, either as the inner or the outer layer, through analytic calculations based on appropriately extended Mie theory. We focus on absorption and scattering of light by bismuth-substituted yttrium iron garnet (Bi:YIG) nanospheres and nanoshells, combined with excitonic materials such as organic-molecule aggregates or two-dimensional transition-metal dichalcogenides, and discuss the hybrid character of the modes emerging from the coupling of the two constituents. We observe the excitation of strong magneto-optic phenomena and explore, in particular, the response and tunability of a magneto-transverse light current, indicative of the photonic Hall effect. We show how interaction between the Bi:YIG and excitonic layers leads to a pair of narrow bands of highly directional scattering, emerging from the aforementioned hybridization, which can be tuned at will by adjusting the geometrical or optical parameters of the system. Our theoretical study introduces optically anisotropic media as promising templates for strong coupling in nanophotonics, offering a means to combine tunable magnetic and optical properties, with potential implications both in the design of all-dielectric photonic devices but also in novel clinical applications.
\end{abstract}

\maketitle


\section{\label{sec:level1}Introduction}

Scattering and absorption by composite multilayered nanoparticles (NPs) have long been at the forefront of interest in nanophotonics, with plasmonic structures providing so far the most prominent and fertile template~\cite{Oldenburg,Teperik,Hao,Tserkezis_jpcm20,Le,Christensen:2015}, aiming to manipulate electromagnetic (EM) fields and generate new, hybrid elements with unique optical properties~\cite{plexcitons_fofang,Schuller,Tserkezis_acsphot5}. Plasmons, i.e., collective oscillations of the conduction-band electrons in metals, are known to exhibit a resonant behavior, tunable through the geometrical and optical parameters of the NP and its environment, triggering impressive optical phenomena, such as huge enhancement and confinement of light in subwavelength volumes~\cite{picocavity}. Nevertheless, high inherent Ohmic losses hinder the widespread use of metals in everyday photonics~\cite{Khurgin}, and focus has recently turned towards high-index dielectrics~\cite{Baranov,Evlyukhin}. In this context, single or composite silicon NPs have been the subject of renewed theoretical and experimental interest, exposing a richness of optical modes, of both electric and magnetic character, generated by oscillating polarization charges and circulating displacement currents inside the particle~\cite{Miemodes, Todisco, Evlyukhin}. In contrast to plasmonic assemblies that usually support negligible magnetic resonances, dielectrics can be fabricated to combine strong magnetic response with low intrinsic losses and enhancement of light comparable to their plasmonic counterparts~\cite{albella_jpcc117,almpanis_josab33}. Moreover, due to their compatibility with existing technologies in microelectronics and the relative ease of fabrication, all-dielectric nanodevices have been proposed as a promising alternative to nanoplasmonics with possible applications in biosensing~\cite{biosensing}, metamaterials~\cite{metasurfaces, metamaterials,Zhu:2017a}, nanoantennas~\cite{nanoantennas_krasnok,nanoantennas_li} and slow light~\cite{raza_ol45}. 

Of particular interest is the case of composite NPs consisting of a dielectric component and an excitonic layer sustained by \textit{J}-aggregates of organic molecules or two-dimensional (2D) transition-metal dichalcogenides (TMDs), operating at or close to the strong coupling regime. Such architectures offer even broader functionality and flexibility in applications, while also providing crucial new insight into the nature and mechanisms governing light-matter interactions. Recently, silicon--\textit{J}-aggregate heterostructures were explored, from both a theoretical and an experimental aspect, as an alternative to plasmon-exciton hybrids termed plexcitons~\cite{plexcitons_fofang,plexcitons_hakala}, revealing the formation of hybrid modes of photonic-excitonic character, termed, in an equivalent manner, Mie-excitons~\cite{Todisco,Mieexcitons,heterostructures,Castellanos,Heilmann}. In particular, since their emergence in literature, it has been envisaged that the complex, magnetic nature of their modes, would eventually allow to externally manipulate them with static magnetic fields~\cite{Mieexcitons} in analogy with active magnetoplasmonics~\cite{lodewijks_nl14}, a feature that has not, however, been explored as yet.

At the same time, composite magnetic NPs with core-shell morphology---usually a magnetic core coated with a biocompatible organic dye---are proposed as suitable building blocks for novel clinical applications in nanomedicine, for a diversity of purposes including imaging, drug delivery and photothermal therapy~\cite{drugdelivery,Pankhurst,Mornet,vollath_2010,hong_2008,chang_2008,lu_2001}. A key advantage of magnetic NPs is their ability to respond to multiple external stimuli (light, magnetic field, temperature, etc) in a non-invasive manner, i.e. without perturbing the biological system. However, the coexistence of light and magnetism in a system characterized by its ability to respond to both gives rise to optical anisotropy, and thus to magneto-optic phenomena, a thorough study of which is required when considering medical applications.
The formulation of Mie scattering by optically anisotropic spheres has already been analytically developed~\cite{Lin} and explicitly performed for plasmon-coated Bi:YIG and magnetite particles of various geometries, exposing strong magnetochirality and a prominent plasmon-driven photonic Hall effect~\cite{PHE, MCD, Yannopapas1, Yannopapas2}---not to be confused with the spin-Hall effect which is purely based on the polarization of light and no magnetic field is needed~\cite{Yin_2013}. In analogy to the classical Hall effect, in its photonic counterpart an incident EM wave propagating through a gyroelectric medium along a direction perpendicular to the applied magnetic field is deflected transversely to both the propagation and the magnetic field direction. Although being essential for understanding the underlying physics and for the design of all-dielectric devices, a thorough investigation of the photonic Hall effect and ways to control it in non-plasmonic core-shell assemblies is still missing. Here, we analyze the photonic Hall effect in composite nanospheres, consisting of a gyroelectric and an excitonic layer, showing that the interaction of Mie resonances with excitonic modes leads to a hybridization manifested through the splitting of the observed magneto-optical response into two narrow bands.

The paper is structured as follows. In Sec.~\ref{sec:level2} we summarize and extend Mie theory for scattering and absorption by coated gyroelectric NPs and nanoshells, and present the scattering cross section formula for the Hall photon current. In Sec.~\ref{sec:level3} we present our theoretical results regarding two specific examples, i.e., a bismuth-substituted yttrium iron garnet (Bi:YIG) nanosphere with an excitonic coating, and an excitonic core coated with a Bi:YIG nanoshell. Our main findings are summarized in the last section of the article.

\section{\label{sec:level2}Theoretical Method}

Let us assume a time-harmonic, monochromatic plane EM wave of angular frequency $\omega$, incident on a gyroelectric sphere of radius $R$ embedded in an infinite homogeneous host medium that is characterized by scalar permittivity and permeability $\epsilon_{2}$ and $\mu_{2}$, in the presence of a static magnetic field. The presence of the magnetic field induces a Lorentz force, which needs to be added in the equations of motion of electrons in the sphere, leading to an anisotropic permittivity tensor~\cite{Wolff}. If the orientation of the magnetic field is along the $z$ axis, the permittivity of the sphere is given by
\begin{gather}\label{perm_tensor}
    \epsilon_1 = \epsilon_z
    \begin{pmatrix}
      \epsilon_r & -i\epsilon_{\kappa} &0 \\
      i\epsilon_{\kappa} & \epsilon_r &0 \\
     0 &0 &1
    \end{pmatrix}
\end{gather}
while the permeability $\mu_1$ is scalar, practically equal to unity in the optical regime~\cite{Landau}. The tensor components are in general complex functions of frequency, taking dispersion and dissipative losses into account while naturally being causal and fulfilling Kramers-Kronig relations.

The fields inside and outside the sphere can be expressed in the basis of vector spherical harmonics. Given the electric field component $\mathbf{E}_0$ of the incoming plane wave, the incident electric field can be written as
\begin{equation}
    \textbf{E}_\textrm{inc} = \textbf{E}_0 \, e^{i\textbf{k}\cdot \textbf{r}} = \sum_{Plm} a^0_{Plm} \, \textbf{F}_{Plm},
\end{equation}
where $P=H,E$ refers to the transverse magnetic (TM) and transverse electric (TE) polarization, $l$ and $m$ are the angular-momentum indices, $\textbf{F}_{Hlm}, \textbf{F}_{Elm}$ are the TM and TE wave functions respectively, explicitly given later on, $\textbf{a}^\textbf{0} = [a_{Hlm}^0 \quad a_{Elm}^0]^T$ with $l \in [1,\infty)$ and $m \in [-l, l]$ is the amplitude of the incident wave (see the Appendix) and $\textbf{a}^\textbf{+} = \textbf{T}  \textbf{a}^\textbf{0}$ is the amplitude of the scattered spherical wave, where $\textbf{T}$ is the scattering matrix. The infinite expansion series describing the fields inside and outside the particle can be truncated in practice at a certain value $l_{\max}$ and in this case the amplitudes are $(n_d \times 1)$ column vectors, where $n_d=l_{\max}(l_{\max}+2)$. In our calculations $l_{\max} = 5$ is adequate to provide converged spectra. It can be shown that $\textbf{T}$ takes the form~\cite{Christofi}

\begin{align}
\mathbf{T} = \mathbf{Z}  (\mathbf{U}+\mathbf{\Lambda}  \mathbf{Z})^{-1}
\end{align}
with
\begin{align}
    \mathbf{Z} = (\mathbf{\Lambda} - \mathbf{\Lambda'})^{-1}  (\mathbf{V}-\mathbf{U}),
\end{align}
where the matrices $\mathbf{\Lambda, \Lambda', V, U}$ are provided in the Appendix.

Having calculated the scattering matrix $\textbf{T}$, the extinction, scattering and absorption cross sections, normalized to the geometric cross section $\pi R^{2}$, are obtained by~\cite{Bohren_Huffman}
\begin{subequations}
\begin{align}
\sigma_\textrm{sc} = \frac{1}{(k_2R)^2 \, \pi |E_0|^2} \sum_{Plm} |a^+_{Plm}|^2 
\end{align}
\begin{align}
\sigma_\textrm{abs} & = -\frac{1}{(k_2R)^2 \, \pi |E_0|^2} \Big\{
  \sum_{Plm} |a^+_{Plm}|^2   \\
& +\textrm{Re} \Big( \sum_{P'l'm'} {a^0_{P'l'm'}}^\dagger  \sum_{Plm} a^+_{Plm} \Big)\Big\}\notag
\end{align}
\begin{align}
& \quad\quad \sigma_\textrm{ext} =\sigma_\textrm{sc} + \sigma_\textrm{abs} =  \\ &-\frac{1}{(k_2R)^2 \, \pi |E_0|^2}
 \textrm{Re} \Big( \sum_{P'l'm'} {a^0_{P'l'm'}}^\dagger  \sum_{Plm} a^+_{Plm} \Big) ,\notag
\end{align}
\end{subequations}

\noindent where $k_2$ is the wavenumber in the host environment. 

\begin{figure}[t!]
\begin{minipage}{0.4\columnwidth}
\begin{itemize}
\item 1: Gyroel. Core
\item 2: Shell
\item 3: Host Medium
\end{itemize}
\end{minipage}
\begin{minipage}{0.5\columnwidth}
 \includegraphics[width=\linewidth]{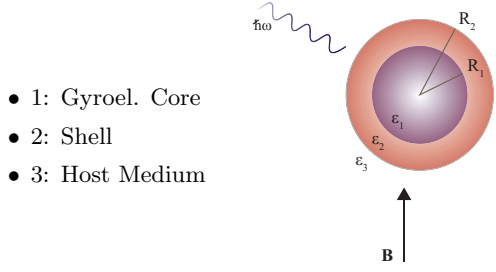}
\end{minipage}
 \caption{\label{fig:gyr_core}
 A nanosphere consisting of a gyroelectric core and an isotropic shell of inner radius $R_1$ and outer $R_2$ illuminated by a plane EM wave in the presence of an external static magnetic field \textbf{B}.}
 \end{figure}
 
We now consider the above gyroelectric sphere of radius $R_1$, coated with a concentric spherical shell of radius $R_2$ and optical parameters $\epsilon_2$ and $\mu_2$---index 3 now refers to parameters of the host medium, as shown in the schematics of Fig.~\ref{fig:gyr_core}.
Boundary conditions at the outer surface $S_2$ of the composite sphere require continuity of the tangential components of the electric and the magnetic field, yielding

\begin{subequations}\label{onion_system}
\begin{gather}
    \mathbf{a^{+} =\tilde{\Lambda} \textbf{a}^{0}  + \tilde{U} \textbf{a}^{0}} \\
 \mathbf{a^{+} =\tilde{\Lambda}' \textbf{a}^{0}  + \tilde{V} \textbf{a}^{0}},
\end{gather}
\end{subequations}
where
\begin{subequations}
\begin{gather}
\mathbf{\tilde{U} = U_A  T + U_B}  \\ 
\mathbf{\tilde{V} = V_A  T + V_B}   
\end{gather}
\end{subequations}
(see Appendix for matrices $\mathbf{\tilde{\Lambda}}$, $\mathbf{\tilde{\Lambda'}}$, $\mathbf{U_A}$, $\mathbf{U_B}$, $\mathbf{V_A}$, $\mathbf{V_B}$).

Equations~(\ref{onion_system}) lead to the following expression for the scattering matrix of the core-shell system:

\begin{gather}
\mathbf{\tilde{T} = \tilde{Z}  \tilde{R}} 
\end{gather}
with
\begin{gather}
\mathbf{\tilde{Z} = (\tilde{U}^{-1} - \tilde{V}^{-1})^{-1}} \\
\mathbf{\tilde{R} = (\tilde{U}^{-1} \tilde{\Lambda} - \tilde{V}^{-1} \tilde{\Lambda}')}.
\end{gather}

\begin{figure}[t!]
\begin{minipage}{0.4\columnwidth}
\begin{itemize}
\item 1: Core
\item 2: Gyroel. Shell
\item 3: Host Medium
\end{itemize}
\end{minipage}
\begin{minipage}{0.55\columnwidth}
 \includegraphics[width=\linewidth]{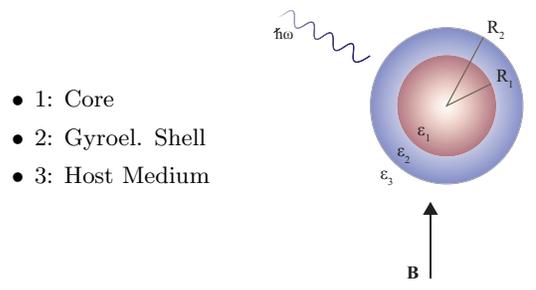}
\end{minipage}
 \caption{\label{fig:gyr_shell}
 A nanosphere of excitonic-gyroelectric core-shell configuration, of inner radius $R_1$ and outer $R_2$, illuminated by a plane EM wave in the presence of an external static magnetic field \textbf{B}.}
\end{figure}

We will now derive the scattering matrix of the inverse core-shell configuration, that is, a gyroelectric shell with a homogeneous medium both inside the cavity and as the host environment, as shown in Fig.~\ref{fig:gyr_shell}.
The wave equation for the electric displacement vector $\mathbf{D}_{2}$ inside the gyroelectric shell can be obtained by substituting Eq.~(\ref{perm_tensor}) into the source-free Maxwell equations, yielding
\begin{equation} 
\nabla \times \nabla \times \Big[ \epsilon_z \epsilon_g^{-1} \textbf{D} (\textbf{r}) \Big] - k_2^2 \, \textbf{D} (\textbf{r}) = 0,
\end{equation}
with $k_2 = \frac{\omega}{c} \sqrt{\epsilon_z\mu_2}$ being the wave number in medium 2 and $c$ the the speed of light in vacuum.

Similarly to the procedure followed in~\cite{Christofi}, one can show now that the fields in the second layer are

 \begin{align}
 \textbf{E} (\textbf{r})  =  \sum_j & b_j  \Big\{ \frac{k_j^2}{k_2^2}  \overline{w}_{00;j} \textbf{J}_{L00}  
 +   \sum_{lm}  \Big[  \frac{k_j^2}{k_2^2} \overline{w}_{lm;j} \textbf{J}_{Llm} \\
 &+ a_{Hlm;j} \textbf{J}_{Hlm} + a_{Elm;j}  \textbf{J}_{Elm} \Big] \Big\} 
 \notag \\
 + \sum_j &  c_j \Big\{ \frac{k_j^2}{k_2^2}  \overline{w}_{00;j} \textbf{H}_{L00} \notag 
 + \sum_{lm}  \Big[ \frac{k_j^2}{k_2^2} \overline{w}_{lm;j} \textbf{H}_{Llm}
 \notag \\
 &+ a_{Hlm;j} \textbf{H}_{Hlm} + a_{Elm;j}  \textbf{H}_{Elm} \Big] \Big\}\notag 
\end{align}  
and
\begin{align}
\textbf{H} (\textbf{r})  =& \sum_j  b_j \frac{k_j^2}{\omega \mu_0 \mu_2}
\sum_{lm} \Big[ a_{Elm;j}  \textbf{J}_{Hlm} 
 -a_{Hlm;j} \textbf{J}_{Elm} \Big] \\
+ \sum_j & c_j \frac{k_j^2}{\omega \mu_0 \mu_2} 
\sum_{lm} \Big[ a_{Elm;j}  \textbf{H}_{Hlm}  
 -a_{Hlm;j} \textbf{H}_{Elm} \Big]   \notag 
\end{align}
with $\textbf{F}=\textbf{J}, \textbf{H}$ satisfying
\begin{gather}
\textbf{F}_{Hlm} (\textbf{r})= f_l(kr), \x  \\
\textbf{F}_{Elm} (\textbf{r})= \frac{i}{k}\nabla  \times f_l(kr), \x  \\
\textbf{F}_{Llm} (\textbf{r})= \frac{1}{k}\nabla [ f_l(kr) Y_{lm} (\hat{\textbf{r}})],
\end{gather} 
representing transverse magnetic, transverse electric and longitudinal wave functions respectively, while $f_l = j_l,h_l^+$ corresponds to either the spherical Bessel or Hankel function of the first kind, $\textbf{X}_{lm}$ are the vector spherical harmonics, $Y_{lm}$ are the ordinary spherical harmonics, and $\hat{\textbf{r}}$ represents the dependence on the polar and azimuthal angle collectively~\cite{Jackson}.

Boundary conditions at the inner ($S_1$) and outer ($S_2$) surface determine the expression for the scattering matrix $\mathbf{T}$:

\begin{gather}
 \mathbf{Z} = (\mathbfcal{L} - \mathbfcal{L} ')^{-1}  (\mathbfcal{V} -\mathbfcal{U} ), \\
 \mathbf{R} = (\mathbfcal{U}  + \mathbfcal{L}    \mathbf{Z})^{-1}, \\
 \mathbf{T} =  \mathbf{Z}   \mathbf{R},
\end{gather} 
where
\begin{subequations}
\begin{gather}
\mathbfcal{U} =  \mathbf{U_1^{S_2}}   \mathbf{M} +  \mathbf{U_2^{S_2}},\\
\mathbfcal{V} =  \mathbf{V_1^{S_2}}   \mathbf{M} +  \mathbf{V_2^{S_2}},\\
\mathbf{M} = \Big( \mathbf{U_1^{S_1}} -  \mathbf{V_1^{S_1}} \Big)^{-1}  \Big( \mathbf{V_2^{S_1}} -  \mathbf{U_2^{S_1}} \Big) .
\end{gather}
\end{subequations}
The matrices entering the above formulas can be found in the Appendix.

 An incident EM wave with linear polarization causes displacement of charge carriers along the direction of the electric field oscillation. The Lorentz force that acts on this movement in the presence of the magnetic field is perpendicular to both the magnetic field and the electric-field polarization ($\textbf{F}_{L} \propto \textbf{v} \times \textbf{B}$), where $\mathbf{v}$ is the velocity of the carriers. 
 In our case, if a $y$-polarized wave propagates along the $x$ axis and the magnetic field is along the $z$ axis, the Lorentz force induces a polarization of charges along the $\hat{\textbf{y}}\times\hat{\textbf{z}}$=$\hat{\textbf{x}}$ axis, corresponding to the photonic Hall effect. As a result, there is a component of light scattered along the $\hat{\textbf{y}}$ direction. It has been shown by Varytis \emph{et~al.}~\cite{PHE} that the scattering cross section of this transverse component, $\sigma_\textrm{Hall}$, along the $\hat{\textbf{y}}$ axis is given by the following exact analytic expression:
\begin{align}
\sigma_\textrm{Hall} &= \frac{1}{\pi |\textbf{E}_0|^2}\frac{2}{(k_2 R)^2}\textrm{Re}\Bigg\{  \notag \\
& \sum_{lm} \Big[ \frac{a^{-m}_l}{l(l+1)}(a^+_{Hlm}a^{+*}_{Elm-1} - a^+_{Elm}a^{+*}_{Hlm-1}) \notag \\
&-\xi^{m-1}_{l-1}(a^+_{Hlm}a^{+*}_{Hl-1m-1} + a^+_{Elm}a^{+*}_{El-1m-1}) \notag \\
&-\xi^{-m-1}_{l-1}(a^+_{Hlm}a^{+*}_{Hl-1m+1} + a^+_{Elm}a^{+*}_{El-1m+1})  \Big] \Bigg\},
\end{align}
where the amplitudes $a_{Plm}^+$ compose the column vector of the scattered wave, with
\begin{equation}\label{coeff}
a_l^m = \frac{1}{2}\sqrt{(l-m)(l+m+1)}
\end{equation}
and
\begin{equation}
\xi_l^m = \frac{1}{2(l+1)}\sqrt{\frac{l(l+2)(l+m+1)(l+m+2)}{(2l+1)(2l+3)}}.
\end{equation}

\section{\label{sec:level3}Results and Discussion}

\begin{figure}[t!]
    \centering
    \includegraphics[width=\columnwidth]{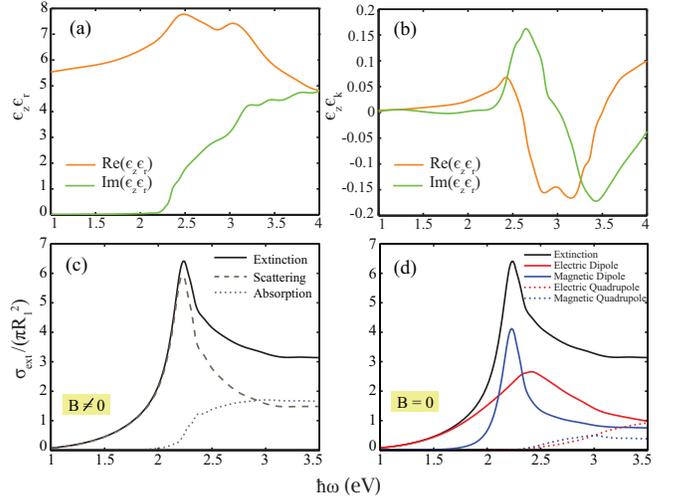}
    \caption{\label{fig:epsilon_BiYIG}
    The real (orange line) and imaginary (green line) part of the (a) diagonal and (b) non-diagonal elements of the permittivity tensor [Eq.~(\ref{perm_tensor})] of Bi:YIG~\cite{BiYIG}. (c) Extinction (black solid line), scattering (grey dashed line) and absorption (grey dotted line) cross sections normalized to the geometric cross section of a Bi:YIG nanosphere of radius $R_1=100$\,nm subjected to a static magnetic field embedded in air. (d) Magnetic (blue) and electric (red) dipolar (solid lines) and quadrupolar (dotted lines) contributions to the extinction cross section (black solid line) for the particle of (c) in the absence of the magnetic field.}
\end{figure}

To design a Mie-excitonic system with strong  photonic Hall effect, comparable to that emerging in plasmonic-gyroelectric structures~\cite{PHE},
we will perform an analytic study of composite core-shell NPs consisting of a gyroelectric and an excitonic layer embedded in air. Unlike plasmonic layers, whose main function is to enhance the near-field and hence any observable effect, the presence of an excitonic layer is expected to lead to a hybridization and an emergence of a tunable double-resonance feature, as we will show below. We assume, to begin with, a plane wave propagating along the $x$ axis, incident on a Bi:YIG sphere of radius $R_1=100$\,nm, while subjected to a static magnetic field oriented along the $z$ axis. Bi:YIG is chosen as a typical gyroelectric high-index dielectric material,
characterized by the experimental optical parameters of~\cite{BiYIG} (measured at saturation) reproduced in Figs.~\ref{fig:epsilon_BiYIG}(a)-(b).

As shown in Fig.~3(c), the extinction and scattering cross sections of this particle, in the visible part of the spectrum, are characterized by a pronounced resonance at $2.24$\,eV, attributed to the magnetic dipolar Mie mode, over a wide but weak electric dipolar background [for the decomposition of the extinction spectrum into its multipolar contributions see Fig.~3(d)], whereas higher-order contributions are almost negligible. This behavior is quite reminiscent of the response of Si NPs~\cite{Miemodes}; what it offers additionally, however, is the non-negligible response to an external magnetic field, contrary to what one might at first anticipate from Figs.~\ref{fig:epsilon_BiYIG}(c) and (d). 
While comparison between the extinction spectra in the presence and absence of the static magnetic field shows that the position and width of the magnetic dipolar resonance is practically unaffected, one should not forget that, first, Bi:YIG reaches saturation at relatively weak magnetic fields~\cite{saturation} and, secondly, any magneto-optic properties, including the photonic Hall effect, are completely eliminated when the field is turned off. The absorption spectrum does not exhibit a Lorentzian-like peak, but a plateau instead, as one could expect from the permittivity data of Figs.~3(a) and (b). The large positive value in the imaginary part of the diagonal elements of the permittivity tensor for energies larger than $2.5$\,eV reveals that this plateau appears most probably due to interband transitions. 

\begin{figure}[t!]
\centering
    \includegraphics[width=\columnwidth]{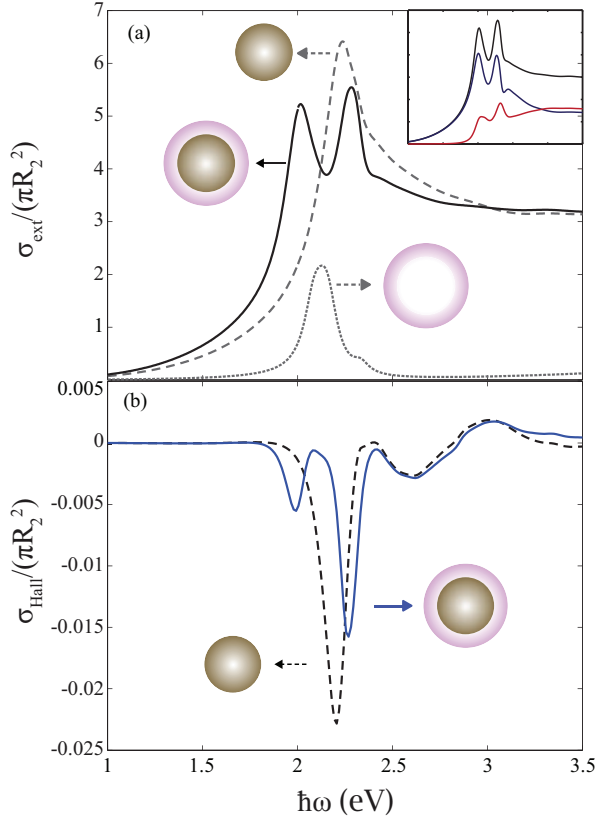}
    \caption{\label{fig:onion}
    (a) Extinction cross section of a Bi:YIG NP of radius $R_1=100$\,nm (grey dashed line), an excitonic shell of inner radius $R_1=100$\,nm and outer $R_2=110$\,nm (grey dotted line), and a core-shell NP consisting of the Bi:YIG core and the excitonic shell (black solid line). The extinction cross section (black line) along with the scattering (dark blue line) and the absorption (dark red line) cross sections of the coupled structure are shown in the inset. (b) Cross section of the magneto-transverse scattered light of the Bi:YIG NP (black dashed line) and the exciton-coated Bi:YIG core (blue solid line), as depicted in the schematics. In all panels air is the host medium.}
\end{figure}

The magneto-optical properties arising in gyrotropic media owe their existence to the non-diagonal components of the permittivity tensor and vanish above the Curie temperature (here $T_C \approx 590$\,K~\cite{Curie_temp}). For the Bi:YIG sphere the non-diagonal elements are large enough to produce magneto-optic phenomena [Fig.~3(b)]. In the present work we shall only be concerned with the photonic Hall effect, but similar conclusions should, in principle, apply to any other manifestations of magneto-optics, such as the Faraday, Kerr, or magnetochiral effects~\cite{Yannopapas2, Varytis_CD, Varytis_FR}.
As displayed in Fig.~\ref{fig:onion}(b) with the black dashed line, a strong component of magneto-transverse scattered light arises at $2.21$\,eV, close to the magnetic dipolar mode, also exhibiting a resonant behavior. 

\begin{figure}[t!]
    \centering
    \includegraphics[width=\columnwidth]{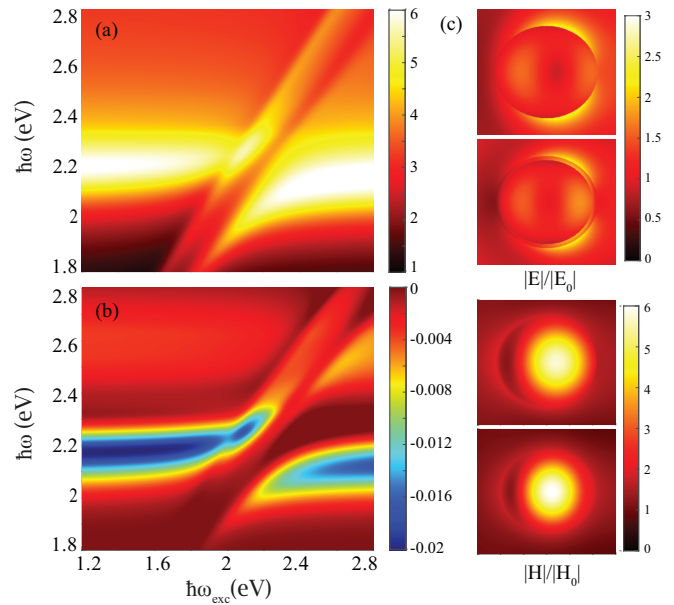}
    \caption{\label{fig:anticrossing}
    (a) Contour plot of the extinction and (b) the magneto-transverse scattered light cross section as a function of the exciton transition energy of the coupled Bi:YIG core-excitonic shell system (c) From top to bottom: Electric near field profile at the energy of the first ($\hbar \omega = 2.01$\,eV) and second mode ($\hbar \omega = 2.28$\,eV) and the magnetic near field profile at the same energies.}
\end{figure}

Let us now consider a composite particle consisting of a spherical Bi:YIG core and a concentric excitonic shell of thickness $R_2-R_1 =10$\,nm.
Such a design can be synthesized relatively easily in the laboratory  and constitutes a flexible platform for engineering the hybrid Mie-excitons~\cite{Mieexcitons}. In practice, an intermediate or outer shell of silica is usually required in fabrication for the protection of the the organic dye~\cite{vollath_2010,hong_2008,chang_2008,lu_2001}. For the excitonic material we use the following generic dielectric function:
 \begin{equation}\label{diel_fun}
\epsilon_\textrm{exc} (\omega) = \epsilon_{\infty} - \frac{f \omega_\textrm{exc}^2}{\omega^2-\omega_\textrm{exc}^2-\mathrm{i}\omega\gamma_\textrm{exc}},
 \end{equation}
where $\omega_{\mathrm{exc}}$ is the excitonic transition frequency, $\gamma_\mathrm{exc}$ the corresponding damping rate, $f$ the oscillator strength and $\epsilon_{\infty}$ the background permittivity. For our calculation we choose $\hbar\omega_\textrm{exc} = 2.12$\,eV, $\hbar\gamma_\textrm{exc}=0.1$\,eV, $f=0.65$ and $\epsilon_{\infty}=3$, values which correspond to an absorption spectrum similar to that of 1,1'-diethyl-2,2'-cyanine iodide (PIC) \textit{J}-aggregates~\cite{struganova_cyanine}. The parameters of the excitonic layer have been chosen so that its resonance frequency lies close to the dipolar magnetic Mie mode of the core. Fig.~\ref{fig:onion}(a) illustrates the extinction spectra of the two constituents individually (grey lines), together with the spectrum resulting from the coupling of the two layers (black solid line). The interaction of the two components leads to the hybridization of their modes in analogy to the formation of bonding and antibonding electron states in molecules, which manifests in the spectra by the emergence of two resonances \cite{Torma}, separated by an anticrossing of width $\hbar \Omega = 0.27$\,eV.
As a result of the addition of the excitonic layer, the pronounced peak of the Hall photon current splits in two peaks  with an energy difference of $0.28$\,eV of only slightly lower intensity [blue line in Fig.~\ref{fig:onion}(b)], following the double peak behavior of the extinction spectrum, already indicating the hybrid nature of the Mie-exciton, since it arises from the coupling of two layers, one of which is an otherwise non-gyroelectric material. The negative cross section values are associated with the direction of choice at which the scattered light is computed, since it is not scattered isotropically.

\begin{figure}[t!]
    \centering
    \includegraphics[width=\columnwidth]{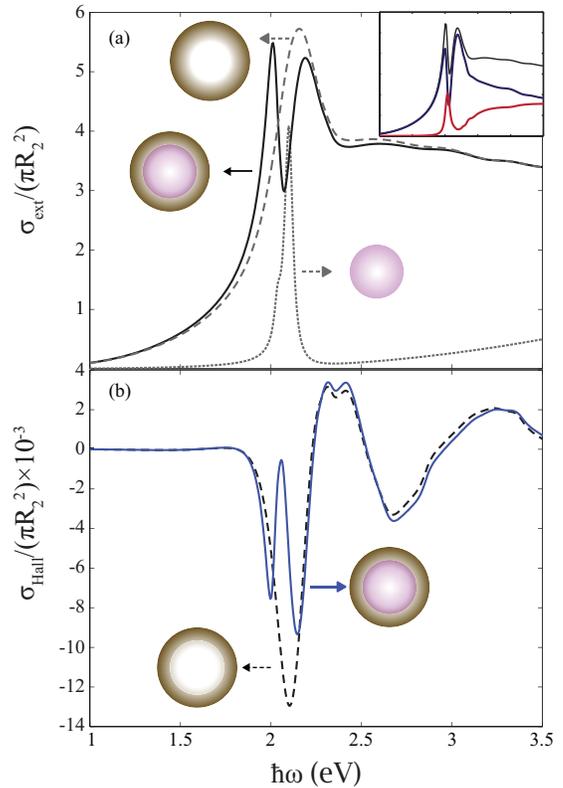}
    \caption{\label{fig:inv_onion}
    (a) Extinction cross section of an excitonic NP of radius $R_1=60$\,nm (grey dotted line), a core-shell NP consisting of a Bi:YIG shell and the excitonic NP of inner radius $R_1=60$\,nm and outer $R_2=110$\,nm, disregarding the excitonic resonance (grey dashed line) and same configuration as taking into account the excitonic transition (black solid line). The extinction cross section (black line) along with the scattering (dark blue line) and the absorption (dark red line) cross sections of the coupled structure are shown in the inset. (b) Cross section of the magneto-transverse scattered light of the Bi:YIG shell (black dashed line) and the Bi:YIG shell-excitonic core NP (blue solid line), as depicted in the schematics illustrations. In all panels air is the host medium.}
\end{figure}

A clear picture of the avoided crossing is provided in Fig.~\ref{fig:anticrossing}(a) by adjusting the values of the exciton transition energy and by extension the detuning of the uncoupled modes. Since it has been demonstrated in Ref.~\cite{enhanced_absorption} that the avoided crossing can also emerge due to enhanced absorption or induced transparency, far from the strong coupling regime, one should also evaluate the absorption spectrum of the setup and the field profile at the energy of the hybrid modes for a more reliable conclusion. Indeed, the strong coupling is verified by the double peak in the absorption spectrum shown in the inset of Fig.~\ref{fig:onion}(a) and the field profiles of Fig.~\ref{fig:anticrossing}(c) of the two coupled modes, which look very similar, but not identical, due to the asymmetry coming from the electric dipolar background. The anticrossing behavior is reproduced in a similar way by the two modes of the Hall photon current, as shown in Fig.~\ref{fig:anticrossing}(b).

In what follows, we invert the arrangement of the two layers and study the photonic Hall effect of an excitonic core-gyroelectric shell configuration of inner and outer radius $R_1=60$\,nm and $R_2=110$\,nm, respectively. Recently, NPs comprising of gyroelectric shells encapsulating non-gyroelectric cores have been proposed as promising templates for hyperthermia applications~\cite{lappas_2019}. It should be noted here that  architectures involving an organic-dye core surrounded by a magnetic shell, challenging as they might be in both fabrication and application in nanobiotechnology, are beneficial in the theoretical search for strongly coupled systems, and they usually provide a clearer physical picture which facilitates understanding of the origin of each spectral feature~\cite{Mieexcitons}. 

\begin{figure}[hbt!]
    \centering
    \includegraphics[width=0.7\columnwidth]{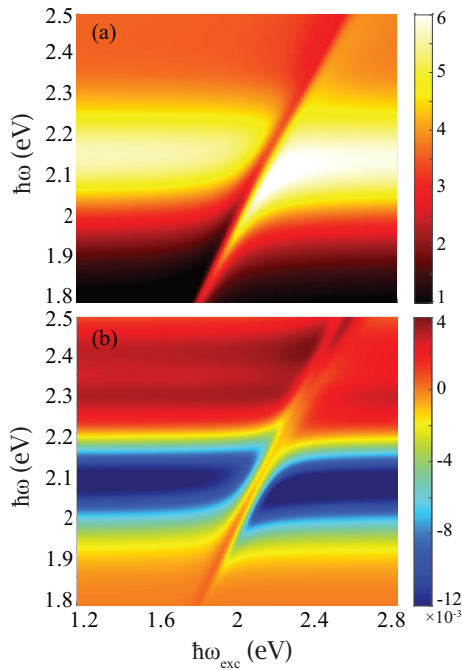}
    \caption{\label{fig:inv_anticrossing}
    (a) Contour plot of the extinction and (b) the magneto-transverse scattered light cross section as a function of the exciton transition energy of the coupled excitonic core-Bi:YIG shell system.}
\end{figure}

The excitonic material is described by the dielectric function of Eq.~(\ref{diel_fun}) with $\hbar\omega_\textrm{exc} = 2.05$\,eV, $\hbar\gamma_\textrm{exc}=0.04$\,eV, $f=0.3$ and $\epsilon_{\infty}=3$. When the excitonic resonance is disregarded, i.e. $f=0$ in the dielectric function of Eq.~(\ref{diel_fun}), the far-field optical response of the Bi:YIG shell [grey dashed line in Fig.~\ref{fig:inv_onion}(a)] does not differ significantly from that of the Bi:YIG sphere of Fig.~3(c), exhibiting a well-defined magnetic Mie resonance at $2.18$\,eV. As depicted in Fig.~\ref{fig:inv_onion}(a), the excitonic core and the gyroelectric shell have been constructed so that the resulting resonances appear at similar energies, leading once again to the double-peak spectrum of the coupled system.
Similarly, the Hall photon current of the gyroelectric nanoshell exhibits resonant spectral features about the Mie mode, as expected. However, Fig.~\ref{fig:inv_onion}(b) shows that it is weaker in comparison to the case of the solid gyroelectric nanosphere. Taking the excitonic resonance into account [$f=0.3$ in Eq.~(\ref{diel_fun})] has similar effect on the $\sigma_\textrm{Hall}$ spectrum as before; namely, the single broad resonance of Bi:YIG at $2.10$\,eV has been split into two narrower ones, which can be important in applications requiring highly directional scattering in a narrow frequency window. In comparison to the inverse configuration of Fig.~\ref{fig:onion}(b), this arrangement exhibits two narrow peaks comparable to each other in both width and magnitude. In this case, the splitting of the two branches in the anticrossing diagrams of Fig.~\ref{fig:inv_anticrossing} both in the extinction spectrum and the Hall cross section is barely discernible and, in addition to the absence of the double peak in the absorption spectrum [inset of Fig.~\ref{fig:inv_onion}(a)], the response resembles a different type of coupling, i.e. what has been termed enhanced absorption~\cite{enhanced_absorption}. Regardless of the coupling characterization, the double peak of Fig.~\ref{fig:inv_onion}(b) reveals the hybrid nature of the two-layered sphere, although not strictly operating in the strong-coupling regime.

\begin{figure}[hbt!]
    \centering
    \includegraphics[width=0.7\columnwidth]{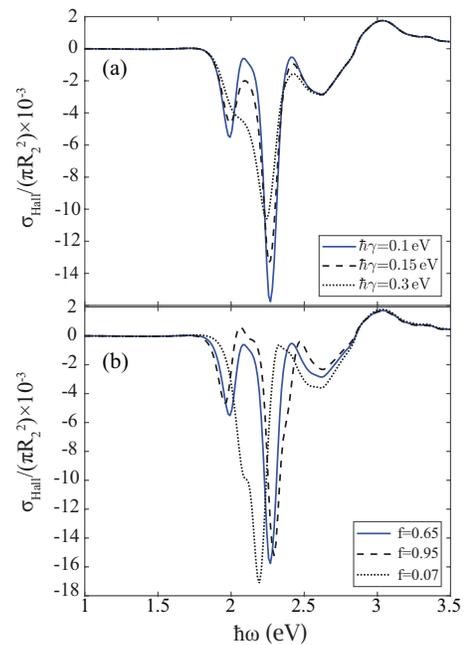}
    \caption{\label{fig:Hall_parameters}
 (a)  Comparison between NPs with different damping rates and (b) oscillator strengths. In both panels the blue solid line corresponds to the cross section of the magneto-transverse scattered light of the Bi:YIG core-excitonic shell NP  of Fig~\ref{fig:onion}.}
\end{figure}

Tunability of these modes is a major advantage. An increase or a decrease in the radius of the composite particle results in redshifting or blueshifting of the Mie modes respectively. It is therefore straightforward that the maximum of the magneto-transverse scattered light shifts accordingly. A rather interesting aspect of the flexibility of such assemblies, especially in strong light-matter interaction studies, is the fact that wider splits of the hybrid modes can be achieved by increasing the oscillator strength of the excitonic dielectric function,  as displayed in Fig.~\ref{fig:Hall_parameters}(b). In particular, in the case of the gyroelectric core--excitonic shell NP, increasing the oscillator strength to $f=0.95$---an increase in the dye thickness would in general have a similar effect~\cite{Mieexcitons}---leads to slightly lower scattering cross section as well as to a wider split up to $0.34$\,eV, which is $0.06$\,eV larger compared to that of Fig.~\ref{fig:onion}(b). 
At the same time, the resonance is not substantially sensitive to the reduction of the oscillator strength, since it has to go below $f=0.07$ for the hybridization to cease completely. On the other hand, the nature of the coupling is drastically altered by the increase of the damping rate. Fig.~\ref{fig:Hall_parameters}(a) shows that for $\hbar\gamma=0.3$\,eV the energy loss mechanism of the excitonic layer is much faster than the energy transfer between the two constituents and, consequently, the composite configuration is no longer under a coupling regime. Alternative routes for higher tunability can be achieved by considering more sophisticated architectures of more layers or of various geometries. Magneto-optical properties of complex plasmon-gyroelectric structures, such as clusters and helices, have already been studied~\cite{Yannopapas1,Yannopapas2}, but a non-plasmonic approach is still missing. A tri-layered system, comprising all three kinds of components (plasmonic, excitonic and magneto-optic) could be a promising route to further enhance the observed effects, although analyzing the hybridization of three different modes in that case might not be straightforward.

\section{\label{sec:level4}Conclusions}
In summary, an analytic method, based on an extended Mie scattering theory, to calculate the Hall photon current for core-shell NPs comprising a gyroelectric layer is provided. Previous work on the photonic Hall effect has been limited to single gyrotropic spheres and metal-coated gyrotropic nanospheres characterized by plasmonic-driven magneto-optical phenomena. Here, strong magnetically induced optic phenomena in dielectrics coupled to excitons are reported by studying the photonic Hall effect in two--layered Bi:YIG-excitonic nanospheres. We show that these composite particles exhibit a rich optical response and a prominent photon Hall current, opening new opportunities for multifunctional dielectric-based photonic platforms, tunable by their geometrical and optical parameters, capable to respond to various external stimuli. Rigorous investigation of the Hall activity is important for clinical applications, especially for techniques in which directionality is a key issue.

\section*{Acknowledgments}
We thank P. Varytis for sharing the interpolated data for the optical parameters of Bi:YIG and C. Wolff for discussions.
P.~E.~S. acknowledges support from an Erasmus+ Scholarship.
%
%
N.~A.~M. is a VILLUM Investigator supported by VILLUM FONDEN (Grant No. 16498) and Independent Research Funding Denmark (Grant No. 7026-00117B).
The Center for Nano Optics is financially supported by the University of Southern Denmark (SDU 2020 funding).

\appendix
\section*{Appendix}
\setcounter{equation}{0}
\renewcommand{\theequation}{A\arabic{equation}}
The amplitudes of the incident field are determined by expanding $\exp(\mathrm{i}\, \textbf{k}\cdot \textbf{r})$ in spherical harmonics~\cite{Jackson}:

\begin{align}
 a_{Elm}^0 =& \frac{4\pi \mathrm{i}^l (-1)^{m+1}}{\sqrt{l(l+1)}}  \notag\\ 
 \times\Big[ &  \{a^m_l Y_{l-(m+1)}(\hat{\textbf{k}}) + a_l^{-m} Y_{l-(m-1)}(\hat{\textbf{k}})\}(\textbf{k} \times \textbf{E}_0)_x + \notag \\
& \mathrm{i}^l  \{a^m_l Y_{l-(m+1)}(\hat{\textbf{k}}) - a_l^{-m} Y_{l-(m-1)}(\hat{\textbf{k}})\}(\textbf{k} \times \textbf{E}_0)_y- \notag \\
& m Y_{l-m}(\hat{\textbf{k}}) (\textbf{k} \times \textbf{E}_0)_z\Big]
\end{align}
and
\begin{align}
 a_{Hlm}^0 =& \frac{4\pi \mathrm{i}^l (-1)^{m+1}}{\sqrt{l(l+1)}}  \notag\\
\times \Big[&   \{a^m_l Y_{l-(m+1)}(\hat{\textbf{k}}) + a_l^{-m} Y_{l-(m-1)}(\hat{\textbf{k}})\} E_{0x}+ \notag \\
& \mathrm{i}^l  \{a^m_l Y_{l-(m+1)}(\hat{\textbf{k}}) - a_l^{-m} Y_{l-(m-1)}(\hat{\textbf{k}})\} E_{0y}- \notag \\
& m Y_{l-m}(\hat{\textbf{k}}) E_{0z} \Big],
\end{align}
where $a_l^m$ is given by Eq.~(\ref{coeff}).

Matrices $\mathbf{\Lambda}, \mathbf{\Lambda'}, \textbf{V}, \textbf{U}$ used in the calculation of the scattering matrix \textbf{T} of a single gyroelectric sphere are given by~\cite{Christofi}

\begin{subequations}
\begin{gather}
 \Lambda_{Plm;P'l'm'} = -\frac{H_{l2}}{J_{l2}} \delta_{ll'} \delta_{mm'}\delta_{PP'} \\ 
 \Lambda'_{Plm;P'l'm'} = -\frac{H'_{l2} }{J'_{l2}}  \delta_{ll'} \delta_{mm'}\delta_{PP'}  \\
 U_{Hlm;j} =  \frac{J_{l;j}}{J_{l2}}a_{Hlm;j}  \\
  U_{Elm;j} = \frac{\mu_2k_j}{\mu_1k_2}\frac{J_{l;j}}{J_{l2}}a_{Elm;j}\\
  V_{Hlm;j} =  \frac{\mu_2}{\mu_1}\frac{J'_{l;j}}{J'_{l2}} a_{Hlm;j}  \\
 V_{Elm;j} =  \frac{k_2}{k_j}\frac{J'_{l;j}}{J'_{l2}}a_{Elm;j} -\frac{ \sqrt{l(l+1)} k_jk_2}{k_1^2}  \frac{J_{l;j}}{J'_{l2}}\overline{w}_{lm;j}. 
\end{gather}
\end{subequations}

In the above formulas we have used the notation $J_{li} = j_l(k_iR)$, $H_{li} = h_l^+(k_iR)$, $J'_{li} = \frac{\partial}{\partial r}[r j_l(k_ir)]|_{r=R}$ and $H'_{li} = \frac{\partial}{\partial r}[ rh_l^+(k_ir)]|_{r=R}$, where $j_l$ and $h_l^+$ are the spherical Bessel and Hankel functions of the first kind, respectively and $k_1 = \frac{\omega}{c}\sqrt{\epsilon_z \mu_1}$ is the wave number of the gyroelectric core. The wave numbers $k_j$ and amplitudes $a_{Plm;j}$ are obtained from the solution of the eigenvalue problem

\begin{equation}
     \sum_{P'l'm'} A_{Plm;P'l'm'} a_{Plm;j} = \frac{k_1^2}{k^2} a_{Plm;j},
\end{equation}
where the subscript $j=1,2,..., 2n_d$ enumerates the eigenvalues and
eigenvectors of matrix \textbf{A} and $b_j$ is a scalar coefficient. Explicit expressions for the matrix elements of A and for $w_{lm;j}$ entering the formulas are provided in Ref.~\cite{Christofi}.

For the scattering matrix of the coated gyroelectric sphere, the following matrices enter into the calculation: 

\begin{subequations}
  \begin{gather}
 \tilde{\Lambda}_{Plm;P'l'm'} = -\frac{J_{l3}}{H_{l3}}  \delta_{ll'} \delta_{mm'}\delta_{PP'} \\
 \tilde{\Lambda}'_{Plm;P'l'm'} = -\frac{ J'_{l3}}{H'_{l3}}  \delta_{ll'} \delta_{mm'}\delta_{PP'} \\
U_{A,Hlm} =  \frac{H_{l2}}{H_{l3}}\delta_{ll'} \delta_{mm'}  \\ 
U_{A,Elm} =  \sqrt{\frac{\mu_3 \epsilon_2}{\mu_2\epsilon_3}}\frac{H_{l2}}{H_{l3}}\delta_{ll'} \delta_{mm'} 
\\
U_{B,Hlm} =  \frac{J_{l2}}{H_{l3}}\delta_{ll'} \delta_{mm'}  \\ 
U_{B,Elm} =  \sqrt{\frac{\mu_3 \epsilon_2}{\mu_2\epsilon_3}}\frac{J_{l2}}{H_{l3}}\delta_{ll'} \delta_{mm'} 
\end{gather}
\begin{gather}
V_{A,Hlm} = \frac{\mu_3}{\mu_2} \frac{H'_{l2}}{H'_{l3}}\delta_{ll'} \delta_{mm'}  \\
V_{A,Elm} =  \frac{k_3}{k_2} \frac{H'_{l2}}{H'_{l3}}\delta_{ll'} \delta_{mm'}
 \\
V_{B,Hlm} = \frac{\mu_3}{\mu_2} \frac{J'_{l2}}{H'_{l3}}\delta_{ll'} \delta_{mm'}  \\
V_{B,Elm} =  \frac{k_3}{k_2} \frac{J'_{l2}}{H'_{l3}}\delta_{ll'} \delta_{mm'}.  
\end{gather}
\end{subequations}

For the scattering matrix of the inverse configuration the matrices used are calculated by

\begin{subequations}
\begin{gather}
U_{1,Hlm}^{S_1} = \frac{J_{l;j}}{J_{l1}} \, a_{Hlm;j} \delta_{ll'} \delta_{mm'} \\
U_{1,Elm}^{S_1} = \frac{k_j\mu_1}{k_1\mu_2} \frac{J_{l;j}}{J_{l1}} \, a_{Elm;j} \delta_{ll'} \delta_{mm'} \\
U_{2,Hlm}^{S_1} = \frac{H_{l;j}}{J_{l1}} \, a_{Hlm;j} \delta_{ll'} \delta_{mm'} \\ 
U_{2,Elm}^{S_1} = \frac{k_j\mu_1}{k_1\mu_2} \frac{H_{l;j}}{J_{l1}} \, a_{Elm;j} \delta_{ll'} \delta_{mm'}
 \end{gather}
 
 \begin{gather}
V_{1,Hlm}^{S_1} = \frac{\mu_1}{\mu_2} \frac{J'_{l;j}}{J'_{l1}} a_{Hlm;j} \\
V_{1,Elm}^{S_1} = \frac{k_1}{k_j} \frac{J'_{l;j}}{J'_{l1}} a_{Elm;j} - \frac{k_jk_1}{k_2^2} \sqrt{l(l+1)} \overline{w}_{lm;j} \frac{J_{l;j}}{J'_{l1}} \\
V_{2,Hlm}^{S_1} = \frac{\mu_1}{\mu_2} \frac{H'_{l;j}}{J'_{l1}} a_{Hlm;j} \\
V_{2,Elm}^{S_1} = \frac{k_1}{k_j} \frac{H'_{l;j}}{J'_{l1}} a_{Elm;j} - \frac{k_jk_1}{k_2^2} \sqrt{l(l+1)} \overline{w}_{lm;j} \frac{H_{l;j}}{J'_{l1}}.
\end{gather}
 \begin{gather}
 \pazocal{L}_{Plm;P'l'm'} = -\frac{H_{l3}}{J_{l3}}  \delta_{ll'} \delta_{mm'}\delta_{PP'} \\
 \pazocal{L}'_{Plm;P'l'm'} = -\frac{ H'_{l3}}{J'_{l3}}  \delta_{ll'} \delta_{mm'}\delta_{PP'} \\
U_{1,Hlm}^{S_2} = \frac{J_{l;j}}{J_{l3}} \, a_{Hlm;j} \delta_{ll'} \delta_{mm'} \\
U_{1,Elm}^{S_2} = \frac{k_j\mu_3}{k_3\mu_2} \frac{J_{l;j}}{J_{l3}} \, a_{Elm;j} \delta_{ll'} \delta_{mm'} \\
U_{2,Hlm}^{S_2} = \frac{H_{l;j}}{J_{l3}} \, a_{Hlm;j} \delta_{ll'} \delta_{mm'} \\
U_{2,Elm}^{S_2} = \frac{k_j\mu_3}{k_3\mu_2} \frac{H_{l;j}}{J_{l3}} \, a_{Elm;j} \delta_{ll'} \delta_{mm'} \\
V_{1,Hlm}^{S_2} = \frac{\mu_3}{\mu_2} \frac{J'_{l;j}}{J'_{l3}} a_{Hlm;j} \\
V_{1,Elm}^{S_2} = \frac{k_3}{k_j} \frac{J'_{l;j}}{J'_{l3}} a_{Elm;j} - \frac{k_jk_3}{k_2^2} \sqrt{l(l+1)} \overline{w}_{lm;j} \frac{J_{l;j}}{J'_{l3}} \\
V_{2,Hlm}^{S_2} = \frac{\mu_3}{\mu_2} \frac{H'_{l;j}}{J'_{l3}} a_{Hlm;j} \\
V_{2,Elm}^{S_2} = \frac{k_3}{k_j} \frac{H'_{l;j}}{J'_{l3}} a_{Elm;j} - \frac{k_jk_3}{k_2^2} \sqrt{l(l+1)} \overline{w}_{lm;j} \frac{H_{l;j}}{J'_{l3}}.
\end{gather}
\end{subequations}



\begin{thebibliography}{56}
\bibitem{Oldenburg} S. J. Oldenburg, R. D. Averitt, S. L. Westcott, and N. J. Halas, Chem. Phys. Lett.  \textbf{288}, 243 (1998).
\bibitem{Teperik} T. V. Teperik, V. V. Popov, and F. J. Garc\'{i}a de Abajo, Phys. Rev. B  \textbf{69}, 155402 (2004).
 \bibitem{Hao} E. Hao and G. C. Schatz, J. Chem. Phys. \textbf{120}, 357 (2004).
 \bibitem{Tserkezis_jpcm20} C. Tserkezis, G. Gantzounis, and N. Stefanou, J. Phys.: Condens. Matter \textbf{20}, 075232 (2008).
\bibitem{Le} F. Le, D. W. Brandl, Y. A. Urzhumov, H. Wang, J. Kundu, N. J. Halas, J. Aizpurua, and P. Nordlander, ACS Nano \textbf{2}, 707 (2008).
\bibitem{Christensen:2015} T. Christensen, A.-P. Jauho, M. Wubs, and N. A. Mortensen, Phys. Rev. B \textbf{91}, 125414 (2015).
 \bibitem{plexcitons_fofang} N. T. Fofang, T.-H. Park, O. Neumann, N. A. Mirin, P. Nordlander, and N. J. Halas, Nano Lett. \textbf{8}, 3481 (2008).
 \bibitem{Schuller} J.~A. Schuller, E.~S. Barnard, W. Cai, Y.~C. Jun, J.~S. White, and M.~L. Brongersma, Nat. Mater. \textbf{9}, 193 (2010).
 \bibitem{Tserkezis_acsphot5} C. Tserkezis, M. Wubs, and N. A. Mortensen, ACS Photonics \textbf{5}, 133 (2018).
\bibitem{picocavity} F. Benz, M. K. Schmidt, A. Dreismann, R. Chikkaraddy, Y. Zhang, A. Demetriadou, C. Carnegie, H. Ohadi, B. D. Nijs, R. Esteban, J. Aizpurua, and J. J. Baumberg, Science \textbf{354}, 726 (2016).
\bibitem{Khurgin} J. B. Khurgin, Nat. Nanotechnol. \textbf{10}, 2 (2015).
\bibitem{Baranov} D. G. Baranov, D. A. Zuev, S. I. Lepeshov, O. V. Kotov, A. E. Krasnok, A. B. Evlyukhin, and B. N. Chichkov, Optica \textbf{4}, 814 (2017).
\bibitem{Evlyukhin} A. B. Evlyukhin, S. M. Novikov, U. Zywietz, R. L. Eriksen, C. Reinhardt, S. I. Bozhevolnyi, and B. N. Chichkov, Nano Lett. \textbf{12}, 3749 (2012).
\bibitem{Miemodes}  A. García-Etxarri, R. Gómez-Medina, L. S. Froufe-Pérez, C. López, L. Chantada, F. Scheffold, J. Aizpurua, M. Nieto-Vesperinas, and J. J. Sáenz,  Opt. Express \textbf{19}, 4815 (2011).
\bibitem{Todisco} F. Todisco, R. Malureanu, C. Wolff, P. A. D. Gonçalves, A. S. Roberts, N. A. Mortensen, and C. Tserkezis, Nanophotonics \textbf{9}, 803 (2020).
\bibitem{albella_jpcc117} P. Albella, M. A. Poyli, M. K. Schmidt, S. A. Maier, F. Moreno, J. J. S\'{a}enz, and J. Aizpurua, J. Phys. Chem. C \textbf{117}, 13573 (2013).
\bibitem{almpanis_josab33} E. Almpanis and N. Papanikolaou, J. Opt. Soc. Am B \textbf{33}, 99 (2016).
 \bibitem{biosensing} O. Yavas, M. Svedendahl, P. Dobosz, V. Sanz, and R. Quidant, Nano Lett. \textbf{17},  4421 (2017).
\bibitem{metasurfaces} S. Jahani and Z. Jacob, Nat. Nanotechnol. \textbf{11}, 23 (2016).
\bibitem{metamaterials}  I. Staude and J. Schilling, Nat. Photonics \textbf{11}, 274 (2017).
\bibitem{Zhu:2017a} X. Zhu, W. Yan, U. Levy, N. A. Mortensen, and A. Kristensen, Sci. Adv. \textbf{3}, e1602487 (2017).
\bibitem{nanoantennas_krasnok} A. E. Krasnok, A. E. Miroshnichenko, P. A. Belov, and Y. S. Kivshar, Opt. Express \textbf{20}, 20599 (2012).
\bibitem{nanoantennas_li} S. V. Li, D. G. Baranov, A. E. Krasnok, and P. A. Belov, Appl. Phys. Lett. \textbf{107}, 171101 (2015).
\bibitem{raza_ol45} S. Raza, Opt. Lett. \textbf{45}, 1260 (2020).
\bibitem{plexcitons_hakala} T. K. Hakala, J. J. Toppari, A. Kuzyk, M. Pettersson, H. Tikkanen, H. Kunttu, and P. Törmä, Phys. Rev. Lett. \textbf{103}, 053602 (2009).
\bibitem{Mieexcitons} C. Tserkezis, P. A. D. Gonçalves, C. Wolff, F. Todisco, K. Busch, and N. A. Mortensen, Phys. Rev. B \textbf{98}, 155439 (2018).
\bibitem{heterostructures} H. Wang, Y. Ke, N. Xu, R. Zhan, Z. Zheng, J. Wen, J. Yan, P. Liu, J. Chen, J. She, Y. Zhang, F. Liu, H. Chen, and S. Deng, Nano Lett. \textbf{16}, 6886 (2016).
\bibitem{Castellanos} G. W. Castellanos, S. Murai, T. Raziman, S. Wang, M. Ramezani, A. G. Curto, and J. {G{\'o}mez Rivas}, ACS Photonics \textbf{7}, 1226 (2020).
\bibitem{Heilmann} R. Heilmann, A. I. Väkeväinen, J.-P. Martikainen, and P. Törmä, Nanophotonics \textbf{9}, 267 (2020).
\bibitem{lodewijks_nl14}K. Lodewijks, N. Maccaferri, T. Pakizeh, R. K. Dumas, I. Zubritskaya, J. Åkerman, P. Vavassori, and A. Dmitriev, Nano Lett. \textbf{14}, 7207 (2014).
\bibitem{drugdelivery} D. K. Kim and J. Dobson, J. Mat. Chem. \textbf{19}, 6294 (2009). 
\bibitem{Pankhurst} Q. A. Pankhurst, J. Connolly, S. K. Jones, and J. Dobson, J. Phys. D: Appl. Phys. \textbf{36}, R167 (2003)
\bibitem{Mornet} S. P. Mornet, S. B. Vasseur, F. Grasset, and E. Duguet, J. Mat. Chem. \textbf{36}, 2161 (2004).
\bibitem{vollath_2010} D. Vollath, Adv. Mater. \textbf{22}, 4410 (2010). 
\bibitem{hong_2008} S. Hong, H. Chen, L. Wang, L. Wang, Spectrochim. Acta A \textbf{70}, 449 (2008).
\bibitem{chang_2008} Q. Chang, L. Zhu, C. Yu, H. Tang, J. Lumin. \textbf{128}, 1890 (2008).
\bibitem{lu_2001} Y. Lu, Y. Yin, B. T. Mayers, and Y. Xia, Nano Lett. \textbf{2}, 183 (2001).
\bibitem{Lin}Z. Lin and S. T. Chui, Phys. Rev. E \textbf{69}, 056614 (2004).
\bibitem{PHE} P. Varytis and N. Stefanou, J. Opt. Soc. Am. B \textbf{33}, 1286 (2016).
\bibitem{MCD} P. Varytis and N. Stefanou, Opt. Commun \textbf{360}, 40 (2016).
\bibitem{Yannopapas1} V. Yannopapas and A. G. Vanakaras, ACS Photonics \textbf{2}, 1030 (2015).
\bibitem{Yannopapas2} V. Yannopapas, Solid State Commun. \textbf{217}, 47 (2015).
\bibitem{Yin_2013} X. Yin, Z. Ye, J. Rho, Y. Wang, and X. Zhang, Science \textbf{339}, 1405 (2013).
\bibitem{Wolff} C. Wolff, R. Rodríguez–Oliveros, and K. Busch, Opt. Express \textbf{21}, 12022 (2013).
\bibitem{Landau} L. D. Landau, E. M. Lifshitz, and L. P. Pitaevskii, \textit{Electrodynamics of Continuous Media}, 2nd ed. (Butterworth-Heinemann, 1984).
\bibitem{Christofi} A. Christofi and N. Stefanou, Int. J. Mod. Phys. B \textbf{28}, 1441012 (2013).
\bibitem{Bohren_Huffman} C. F. Bohren and D. R. Huffman, \textit{Absorption and Scattering of Light by Small Particles} (Wiley, 1983).
\bibitem{Jackson} J. D. Jackson, \textit{Classical Electrodynamics}, 3d ed. (Wiley, 1999).
\bibitem{BiYIG} V. Doormann, J. P. Krumme, and H. Lenz, J. Appl. Phys. \textbf{68}, 3544 (1990).
\bibitem{saturation} T. Kim, S. Nasu, and M. Shima, J. Nanopart. Res. \textbf{9}, 737 (2006).
\bibitem{Curie_temp} P. Hansen, K. Witter, and W. Tolksdorf, Phys. Rev. B \textbf{27}, 6608 (1983).
\bibitem{Varytis_CD} P. Varytis, N. Stefanou, A. Christofi, and N. Papanikolaou, J. Opt. Soc. Am. B \textbf{32}, 1063 (2015).
\bibitem{Varytis_FR} P. Varytis, P. A. Pantazopoulos, and N. Stefanou, Phys. Rev. B \textbf{93}, 214423 (2016).
\bibitem{struganova_cyanine} I. Struganova, J. Phys. Chem. A \textbf{104}, 9670 (2000). 
\bibitem{Torma} P. Törmä and W. L. Barnes, Rep. Prog. Phys. \textbf{78}, 013901 (2014).
\bibitem{enhanced_absorption} T. J. Antosiewicz, S. P. Apell, and T. Shegai, ACS Photonics \textbf{1}, 454 (2014).
\bibitem{lappas_2019} A. Lappas, G. Antonaropoulos, K. Brintakis, M. Vasilakaki, K. N. Trohidou, V. Iannotti, G. Ausanio, A. Kostopoulou, M. Abeykoon, I. K. Robinson, and E. S. Bozin, Phys. Rev. X \textbf{9}, 041044 (2019).
\end{thebibliography}
\end{document}